# Chemical aging of large-scale randomly rough frictional contacts

**Authors:**

Zhuohan Li[1], Lars Pastewka[2], Izabela Szlufarska[1, *]

**Affiliations:**

[1]Department of Materials Science and Engineering, University of Wisconsin-Madison, Madison, 53706-1595, USA

[2] Department of Microsystems Engineering, University of Freiburg, Georges-Köhler-Allee 103, 79110 Freiburg, Germany

* szlufarska@wisc.edu

**Abstract**

It has been shown that contact aging due to chemical reactions in single asperity contacts can have a significant effect on friction. However, it is currently unknown how chemically-induced contact aging of friction depends on roughness that is typically encountered in macroscopic rough contacts. Here, we develop an approach that brings together a kinetic Monte Carlo model of chemical aging with a contact mechanics model of rough surfaces based on the boundary element method to determine the magnitude of chemical aging in silica/silica contacts with random roughness. Our multi-scale model predicts that chemical aging for randomly rough contacts has a logarithmic dependence on time. It also shows that friction aging switches from a linear to a non-linear dependence on the applied load as the load increase. We discover that surface roughness affects the aging behavior primarily by modifying the real contact area and the local contact pressure, whereas the effect of contact morphology is relatively small. Our results demonstrate how understanding of chemical aging can be translated from studies of single asperity contacts to macroscopic rough contacts.

## I. INTRODUCTION

The phenomenon of contact aging has been known since the pioneering work of Dieterich who found that static friction in some macroscopic contacts increases logarithmically with time for which the contact was held still prior to sliding [1]. Since then, a number of geological materials have been shown to undergo macroscopic aging, [2] a process which is considered to be critical for controlling earthquake nucleation and recurrence. More recently, contact aging has been also observed in nanoscale contacts for a number of material systems, including a silica tip on a silica substrate [3], a silicon tip on a diamond substrate [4], a silicon tip on a highly oriented pyrolytic graphite (HOPG) substrate [5], and an antimony nanoparticle on a HOPG substrate [6,7]. Contact aging can be qualitatively captured by so-called rate- and state- friction laws [8–11]. These laws assume some kind of evolution of the contact states, which leads to change in the friction force, but the physical nature of this evolution is not well understood and has been a subject of debate. There are two general hypotheses for the mechanisms underlying contact aging. One hypothesis assumes evolution of the *"contact quantity"*, which means that contact area grows over time due to plastic creep [12,13], dissolution and precipitation of material at the contact periphery [14,15], or atomic attrition [16]. The second hypothesis explains aging as the process of increasing the *"contact quality"*, which means that the contact becomes mechanically stronger due to chemical bonding [3], capillary interactions [17], evolution from an incommensurate contact to a commensurate contact [4], or local pinning into deeper energy minima [18]. The increase in contact area has been observed in macroscopic slide-hold-slide experiments [12–14,19], but it has been challenging to exclude the evolution of contact quality due to the limitations of macroscopic experiments [20]. In contrast, nanoscale experiments and numerical simulations provide ideal contact conditions for exploring specific aging mechanism and for isolating their contributions. For instance, the use of small contact pressure (~ 1 GPa) relative to the material hardness (~10 GPa for quartz/amorphous silica [21]) in atomic force microscope (AFM) experiments prevents plastic deformation. This approach has been used in nanoscale $SiO_2$-$SiO_2$ contacts to provide the first evidence of chemical contributions (interfacial covalent bonds) to contact aging [3,22]. The authors found that in $SiO_2$-$SiO_2$ single asperity nanoscale contacts there is a logarithmic increase of friction with hold time [3] and a linear dependence of aging on the normal load [22]. The corresponding aging mechanism was revealed by a kinetic Monte Carlo (kMC) model [23], which showed that the logarithmic time dependence of aging in single asperity contacts can arise purely from the formation of siloxane bonds at the hydroxylated $SiO_2$-$SiO_2$ interface [24], based on the following reaction

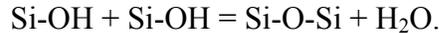

$$\text{Si-OH} + \text{Si-OH} = \text{Si-O-Si} + H_2O.$$

More recently, new developments have been made to the atomistically-informed kMC model for single asperity chemical aging to include normal load dependence [25], and the model showed a quantitative agreement with results of AFM experiments on load dependence of aging [22]. Chemical aging has also been observed in other materials systems, such as zinc dialkyldithiophosphates [26] and diamond-like-carbon [27], which points to the universality of chemical aging of frictional interfaces.

While chemical aging has been demonstrated to be important for single asperity contacts, one needs to ask how significant is the contribution from interfacial chemical bonding to contact aging of large-scale rough contacts and how it depends on the details of the surface statistics [28–31] (e.g., the real contact area, the local pressure distribution, and contact morphology).

Here, we extend our atomistically-informed kMC model for chemical aging [23,25] to investigate rough contacts and to determine how different aspects and parameters of surface roughness affect contact aging. Specifically, the kMC model is combined with boundary element method (BEM) [32–35] to investigate time-dependent bonding of $SiO_2$-$SiO_2$ elastic randomly rough contacts. This multi-scale model allows us to determine how chemical reactions and large-scale deformation are coupled together to produce time-dependent friction. The results are compared to previously reported experimental measurements on macroscopic rough contacts.

## II. MODEL
### A. Boundary element method (BEM) for rough surfaces

Surface roughness can be characterized by the power spectral density (PSD) $C(\mathbf{q})$ [36], which is the Fourier transform of the spatial autocorrelation function of the map of surface heights $h(x, y)$. Here, the wavevector lies within the surface plane, $\mathbf{q}$ = ($q_x$, $q_y$). For surfaces with isotropic statistics, which is assumed in our simulations, $C(\mathbf{q})$ depends only on the magnitude of the wavevector $q = |\mathbf{q}| = 2\pi/\lambda$, where $\lambda$ is the wavelength. Many physical surfaces [37–39] show characteristics of self-affine fractal surface with roughness extending over several orders of magnitude in length [40]. For a self-affine fractal surface, the root-mean-square (rms) height over lateral length $\ell$ scales as $\ell^H$, where $H$ is the Hurst exponent with value between 0 and 1. An implication of this is that $C(\mathbf{q})$ of a self-affine fractal surface has a power-law dependence on the wavevector $q$. Surfaces are often not fractal over all scales but only within a range of wavelengths $\lambda_r$ to $\lambda_s$, where $\lambda_s$ can be close to the atomic scale [41]. An idealized PSD has power-law scaling from the large wavevector cut-off $q_s = 2\pi/\lambda_s$ to the roll-off wavevector $q_r = 2\pi/\lambda_r$ [40]. It has constant power for smaller wavevectors down to $q_L = 2\pi/\lambda_L$, where $\lambda_L$ is the long wavelength cut-off, often the system size. This idealized $C(\mathbf{q})$ can be expressed as

$$C(\mathbf{q}) = C_0 \begin{cases} q_r^{-2-2H}, & \text{if } q_L < |\mathbf{q}| < q_r \\ |\mathbf{q}|^{-2-2H}, & \text{if } q_r < |\mathbf{q}| < q_s \end{cases} \quad (1)$$

Here $C_0$ depends on H, $q_s$, $q_r$, $q_L$ and one additional parameters of the surface roughness, such as the rms slope $h'_{rms} = \sqrt{\langle |\nabla h|^2 \rangle}$ or rms height $h'_{rms} = \sqrt{\langle |h|^2 \rangle}$. The angle brackets $\langle \cdot \rangle$ indicate an average over the full surface.

A realization of a surface $h(x, y)$ can be generated from $C(\mathbf{q})$ using a Fourier filtering algorithm [29,36]. As an example, Figure 1 shows the surface topography of a self-affine fractal surface generated with roughness parameters of $h'_{rms} = 0.1$, $H = 0.8$, $\zeta = q_s/q_L = 150$, $\eta = q_r/q_L = 3$. The long wavelength cut-off $\lambda_L$ is chosen to be equal to the system size, which is $2500a$, where $a = \sqrt{1/\rho_{OH}} \approx 0.45$ nm is the pixel size. The density of OH groups on the surface, $\rho_{OH} = 4.9$ OH/nm$^2$, is the density obtained for a fully hydroxylated silica surface under standard temperature and pressure conditions [42]. Thus, each square pixel represents one OH group. Here, the nominal contact area $A_0 = \lambda_L^2 \approx 1.3$ μm$^2$ is much larger than the typical area of an AFM single asperity contact (~$10^2$ nm$^2$ [22]). Increasing the system size further would increase the computational cost but the qualitative results presented in this paper do not change and thus we will focus our analysis on surfaces with the length scale on the order of a micrometer.

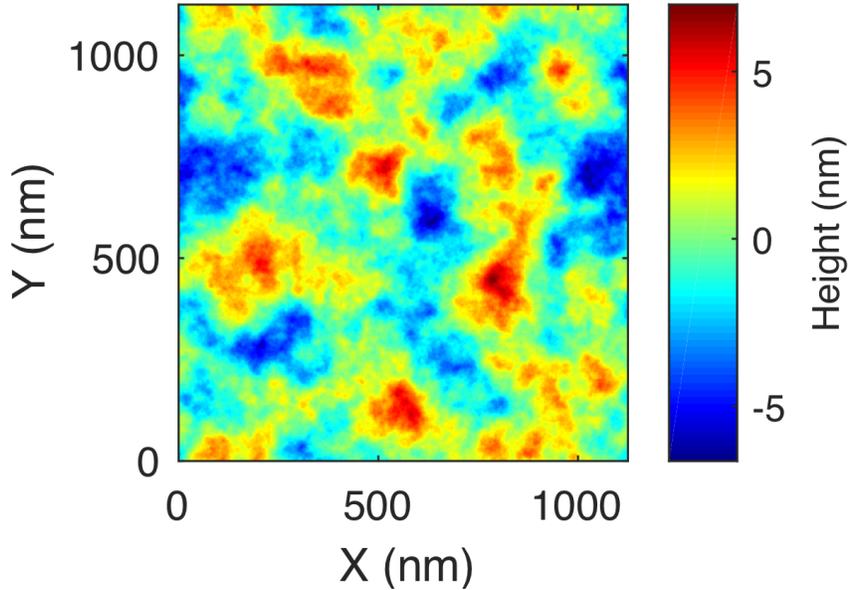

FIG. 1 (color online) Surface topography of a self-affine fractal surface with roughness parameters of $h'_{rms} = 0.1$, $H = 0.8$, $\zeta = 150$, $\eta = 3$.

Chemical aging occurs during stationary contact of two surfaces. Here, we use BEM simulations to obtain the real contact area, as well as the local contact pressure distribution across the interface under applied normal loads. We also assume that the contact is a non-adhesive and frictionless and it occurs between two elastic randomly rough surfaces described by self-affine fractals [36]. In a linear elastic approximation, this type of contact can be mapped onto the contact between a flat elastic substrate and a rough rigid surface [43]. Plastic deformation and creep are ignored in order to study the effect of chemical bonding alone, but we will discuss the possible impact of plastic deformation on our results at the end of this paper. We will also neglect adhesive effects arising from the formation of siloxane bonds. Adhesion affects the local pressure distributions especially at the periphery of the contacts [44], where tensile stress exists. For hard materials, macroscopic adhesion is in practice manifested only in small contacts, such as single asperity contacts [22], or for exceptionally smooth macroscopically rough surfaces ($h'_{rms} \lesssim 10^{-3}$) [44].

In our BEM calculations, we use the reciprocal space formulation of surface Green's function for a linear elastic, isotropic solid of infinite thickness [35,45] with an effective Young's modulus (contact modulus [46]) $E^* = [2(1 − \nu^2)/E]^{-1}$ = 35.3 GPa. Young's modulus $E$ = 69.0 GPa and Poisson ratio $\nu$ = 0.19 represent the $SiO_2$-$SiO_2$ interface [22]. Periodic boundary conditions are applied in the plane of the surface. The elastic problem is then solved using an efficient Fast-Fourier transform technique [33]. The condition at contact are impenetrable hard-walls [32]. We verified that our model reproduces the known relationship between the real contact area ($A$) and the normal load ($F_N$) at small normal loads [28,30,31]

$$A = \frac{\kappa F_N}{h'_{rms} E^*} \tag{2}$$

where $\kappa$ is dimensionless proportionality coefficient with a typical value of 2. Our BEM simulations show a linear area-load relationship at small normal loads corresponding to fractional contact areas $A/A_0$< 0.1 with apparent surface area $A_0 = \lambda_L^2$ and $\kappa$ = 2.1 for the rough surface shown in Fig. 1. A small deviation from the linear relationship is observed at high normal loads, which agrees with previous results of numerical simulations [28,30,31] as shown in supplemental material [47]. Surfaces with other roughness parameters considered in this paper (see Table 1 and Fig. 3) also show a similar trend with $\kappa \approx$ 2-2.3.

### B. kinetic Monte Carlo model

The kMC model used in the current study is the same as the one used for single asperities and the details can be found in Ref. [25]. A similar model, referred to as "mechao-kinetic model", was also previously developed by Barel *et al.* [48]. The model considered bond formation and rupture, and the effect of these processes on friction. The model was applied to simulate aging due to interfacial capillary bridges [49].

Since the details of our kMC model have been reported before [25], here we focus on the key assumptions of the model. First, we assume there is an intrinsic energy barrier distribution to bond formation $E_{b,form,i}(x,y)$ that fluctuates across the surface and the actual energy barriers $E_{b,form}(x,y)$ are affected by the local contact pressure $P(x,y)$ following the Eyring relationship [22,23],

$$E_{b,form}(x,y) = E_{b,form,i}(x,y) − \Delta V P(x,y) \tag{3}$$

where $\Delta V$ is the activation volume. It should be noted that the Eyring relationship is sometimes used to model the effect of Hertzian pressure (mean pressure inside the contact, $\bar{P} =$

$F_N/A_0$) [22,50–52], whereas in our model we explicitly consider the effect from local pressure on each reaction site [23]. Previous theoretical work on single asperity contacts found that the specific shape of the energy barrier distribution does not affect the qualitative kMC results [23]. Irrespectively of the underlying distribution there is always a set of physically-justifiable parameters [25] that can produce a quantitative agreement with AFM experiments [22]. One of the distributions considered in Ref. [23] was determined directly from molecular dynamics (MD) simulations, which is the distribution that we use for $E_{b,form,i}$ in the current study.

The second factor considered here is the interaction between siloxane bonds when they try to form at neighboring reaction sites, which is mediated by elastic deformation of tetrahedra in the bulk silica [23]. The interaction results in the change in the reaction energy barrier $\Delta E_{b,form}$ which is determined following the same expression as used in Refs. [23,25]

$$\Delta E_{b,form} = I_b * (\phi - \varepsilon) \tag{4}$$

where $I_b$ is the maximum interaction value with $\varepsilon = 0$, and $\phi$ is a random number drawn from a uniform distribution between 0 and 1 that is generated for each neighboring reaction site pairs. The variable $\varepsilon = 0.1$ represents a system whose elastic interaction is strongly biased to positive values, consistently with what is reported in Ref. [23]. In our simulation, this interaction is considered only between 8 nearest neighbors. By combining Eq. (3) and Eq. (4) and taking into account the fact that the energy barrier distribution $E_{b,form}$ depends on time, the final expression for the energy barrier to bond formation takes the following form,

$$E_{b,form}(x,y,t) = E_{b,form,i}(x,y) - \Delta VP(x,y) + \sum_{n=1}^{8} \gamma_n(t)\Delta E_{b,form,n} \tag{5}$$

where $\Delta E_{b,form,n}$ is the change in energy barrier to bond formation due to the interaction with the $n^{th}$ neighbor reaction site. $\gamma_n = 1$ if bond at $n^{th}$ neighbor reaction site has already formed, and otherwise $\gamma_n = 0$.

In our model, we also allow siloxane bonds to break during the aging process. It has been reported that water molecules can weaken siloxane bonds by hydrolysis reaction when the bonds are strained (which is the case for the interfacial siloxane bridges under normal stress) [53–55]. Density functional theory calculations have shown that the range of energy barriers to break an interfacial bond is much narrower than the range of energy barriers to bond formation [see Fig. S3(b) in Ref. [23]]. In the current model, the energy barrier to bond breaking $E_{b,break}$ is assumed to be a narrow Gaussian distribution centered at 1.1 eV [25,56].

The average bond rupture force $\langle f \rangle$ is a function of the energy barrier to bond breaking, the temperature, and the pulling velocity [57]. In our model, we assume a constant temperature and a constant pulling velocity of the experiment. Besides, the narrow energy barrier to bond breaking used in our model ensures that on the average the kinetics of bond breaking is quite uniform across the entire interface. Therefore here, we use the average bond rupture force $\langle f \rangle$ = 3.0 nN [58], which is between the theoretical value (4.4-6.6 nN) [59] obtained from *ab initio* MD simulation of the siloxane bonds with high pulling velocity and the mean rupture force (~1.3 nN) [60] measured for the siloxane bonds in the polymer backbone of polydimethylsiloxane by AFM single molecule experiments. The actual bond rupture force $\langle f \rangle$ may be different, but it should only be affected by a factor of less than ~2. The reduction of bond rupture force (relevant for bond rupturing at the onset of sliding) due to the elastic interaction [61] is not included because it has been previously reported that this effect is relatively weak [25] This is because strong elastic interaction prevents new bonds from forming at sites that are neighboring to an existing bond, and thus the rupture force of formed bonds will not be affected due to the limited number of interfacial bonds on neighboring sites. In the case of bonds with a weak elastic interaction, bonds may form at the neighboring reaction sites, however the reduction of rupture force is also small due to fact that the elastic energy is small.

In experiments, the amount of aging is reflected in the value of friction drop $\Delta F$, which is the difference between the static and kinetic friction after a certain hold time. The hold time is defined as the contact time during the stationary contact with a constant normal load and with or without shear stress. In our model, we assume zero shear stress during the stationary contacts and all siloxane bonds are broken at the onset of sliding and hence $\Delta F = N\langle f \rangle$, where $N$ is the number of siloxane bonds formed across the interface. This is supported by MD simulations that show that the friction force is directly proportional to the number of bonds formed across an interface [62]. If $\Delta F$ is divided by the applied normal load $F_N$, one can determine the drop in the coefficient of friction $\Delta \mu = \Delta F / F_N$.

### C. Hierarchical coupling between BEM and kMC

BEM calculations and kMC simulations are coupled hierarchically. Once the contact morphologies and local pressure distributions are obtained from BEM calculations, these values are used as inputs for the subsequent kMC simulations. It is assumed that only those reaction sites (OH groups) within the contact regions (with a positive local pressure) can react and form the siloxane bonds across the interface.

All the kMC simulations shown in the following sections are conducted at room temperature ($T = 300$ K) and the attempt frequencies for siloxane bond formation and breaking during aging are set as $10^{13}$ Hz, which is the typical atomic vibration frequency. Also, all the results are averaged over 10 separate simulations that use the same realization of the rough surface (and hence the same pressure distribution from the BEM calculations), but with different random numbers in the kMC simulations. The error bars of simulation data are not shown except in Fig. 6, because the size of the error bars is comparable to the symbol size.

### III. RESULTS
#### A. Time and load dependence of chemical aging

We first adapt our kMC model to investigate time and load dependence of chemical aging in large randomly rough contacts. The initial energy barrier distributions used in the simulations are shown in Fig. 2 (a). The distribution of $E_{b,form,i}$ (blue curve) is of the same type as the energy barrier distribution shown in Fig. 4 (c) of Ref. [23], but here the range of energy barriers is scaled to be 0.8-1.4 eV: The minimum value $E_{b,form,i} = 0.8$ eV is selected so that the logarithmic time-dependence of aging starts at $t \sim 1$ s (see Fig. 2 (b)), which is the typical hold time for the beginning of the logarithmic aging behavior observed in macroscopic experiments [1,2,63]. The maximum value $E_{b,form,i} = 1.4$ eV is the same as the value used in Ref. [23]. The initial energy barrier to bond formation $E_{b,form}(t = 0)$ of each reaction site follows the distribution shown in red in Fig. 2(a), which includes the effect of local pressure and is related to $E_{b,form,i}$ through Eq. (3). For randomly rough contacts, the distribution of local contact pressures $P$ is approximately independent of the applied normal load $F_N$ for sufficiently small loads when normalized by the real contact area [30,40]. Therefore, the distribution of $E_{b,form}(t = 0)$, which depends on the local contact pressure $P$, also remains almost the same as the normal load $F_N$ increases (see Supplemental Material [47]).

In addition, we assume a relatively large elastic interaction $I_b$ = 0.5 eV between neighboring bonds (for single asperity contacts $I_b$ was assumed to be equal 0.1 eV [25]). This is reasonable because earlier density functional theory calculations revealed that $I_b$ increases with an increasing indentation (contact penetration) depth [23], and the indentation depth of macroscopic rough contact should be larger than the corresponding value in a single asperity contact. Simulations reported in Ref. [25] have shown that for a single asperity contact, $\Delta F$ decreases as $I_b$ increases, but $\Delta F$ saturates when $I_b \gtrsim$ 0.5 eV. Therefore, here we assumed the value of $I_b$ that corresponds to the onset of saturation in $\Delta F$.

The results of kMC simulation for the surface shown in Fig. 1 at normal loads $F_N$ = 20-500 µN are shown in Figs. 2(b)-(d). This range of loads correspond to the fraction of real contact area $A/A_0$ = 0.01-0.22. Our kMC model predicts that the friction drop $\Delta F$ depends logarithmically on time [Fig. 2 (b)] and linearly on the normal load [Fig. 2 (c)]. These trends for large randomly rough contacts are qualitatively similar to those observed for single asperity contacts [3,22,23,25]. We found robust logarithmic dependence between times $t$ = 1s to $10^5$ s, consistent with macroscopic experiments [1,11,63].

We found that there is a small deviation from the linear dependence of $\Delta F$ on normal load for the case of sufficiently large loads, which results in the decrease in the drop in the coefficient of friction $\Delta \mu = \Delta F/F_N$ as shown in Fig. 2 (c). This deviation is reminiscent of similar deviations observed for the area-load dependence of randomly rough contacts obtained from BEM calculation discussed in section II. Specifically, the real contact area is linear with the applied load for smaller loads (for smaller real contact areas) but deviation from linearity is observed at larger load. To demonstrate that the deviation from linearity of $\Delta F$ vs. load relationship originates from the dependence of load on the real contact area, in Fig. 2 (d) we plot $\Delta F$ (and $\Delta \mu$) as a function of $A/A_0$. It is clear that $\Delta F$ is linear with $A/A_0$ up to the highest real contact area considered here. This indicates that the real contact area $A$ is an important factor for the chemical aging of rough contacts. It is interesting to compare this observation with the result of single asperity chemical aging reported in Refs. [22,25], where contact area was also found to play an important role due to the fact that the number of initial reaction sites available for siloxane bond formation is determined by the contact area. Our single asperity chemical aging model [25] revealed that the friction drop $\Delta F$ scales with contact area for a wide range of hold times, even if multiple effects that in principle affect the area dependence (e.g. pressure dependent energy barriers) are considered. The results shown here further indicate that the dominant role of the real contact area in chemical aging is still true for large-scale randomly rough contacts, which consist of multiple smaller contact patches with different contact areas and local pressures. Besides, $\Delta \mu$ stays almost constant as a function of $A/A_0$, especially at small normal loads, which implies that it is possible to bridge the gap of aging behavior between nanoscale single asperity contacts to macroscale rough contacts simply by scaling the contact area. However, it is still unclear how the roughness of the surface affects this scaling, which question will be investigated next. Also, it should be noted that the value of $\Delta \mu$ shown here is orders of magnitude higher than typical value measured in macroscopic experiments [1]. Likely reasons for this deviation from experiments will be discussed in section IV.

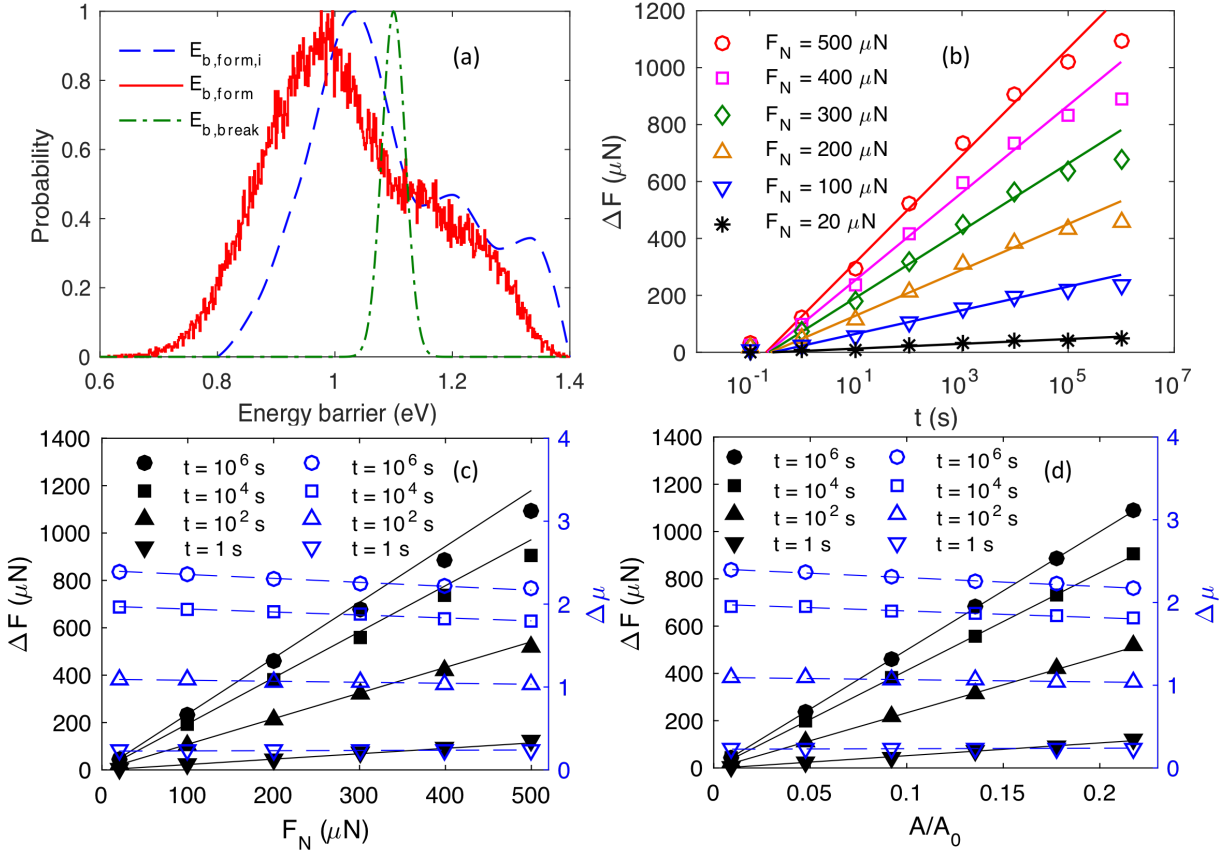

*FIG. 2 (color online) Results of kMC simulations for a randomly rough surface shown in Fig. 1. (a) Distribution of energy barriers. Blue dashed, red solid, and green dash-dot curve represents distributions of intrinsic energy barrier to bond formation $E_{b,form,i}$, initial energy barrier to bond formation $E_{b,form}(t=0)$ under normal load $F_N = 20$ μN, and energy barrier to bond breaking $E_{b,break}$, respectively. All the distributions are scaled so that the maximum is 1 for an easy comparison. (b) Friction drop $\Delta F$ as a function of hold time t for multiple normal loads $F_N$. $\Delta F$ can be divided by $\langle f \rangle = 3.0$ nN in order to convert it to the number of bonds formed across the interface. (c) Load and (d) contact area dependence of $\Delta F$ and $\Delta \mu$ at different hold times. Fraction of real contact area $A/A_0$ is the ratio of the real to the nominal contact area. All results shown in (b)-(d) are averaged over 10 different simulations performed for the same conditions (error bars are smaller than the symbol size).*

### B. Effect of surface roughness parameters on chemical aging

In order to determine the roughness effects on chemical aging, we performed a series of kMC simulations for surfaces with different roughness parameters, as shown in Fig. 3. The roughness parameters varied here are rms slope $h'_{rms}$, Hurst exponent H, and the parameters characterizing the bandwidth of fractal scaling, $\zeta = q_s/q_L$ and $\eta = q_r/q_L$. The ranges of these parameters are listed in Table 1. All the simulation results are presented as the drop in the friction coefficient $\Delta \mu = \Delta F/F_N$ instead of the drop in the friction force $\Delta F$ in order to be consistent with most published papers on macroscopic aging. These papers typically report $\Delta \mu$ along with the slope $\beta = \Delta \mu / \Delta \log(t)$ [2,20,64–66]. As a representative example, we plot the results for $F_N = 20$ μN

(normal pressure $P_N = F_N/A_0 = 16$ MPa). It was demonstrated that $\Delta\mu$ remains almost constant with increasing $F_N$ as shown in Fig. 2 (c), and thus a relatively small normal load is used here to reduce the simulation time.

Fig. 3 (a) shows that the regime of logarithmic dependence of $\Delta\mu$ on hold time exists for all randomly rough surfaces considered here, but there are three distinct regimes that strongly depend on the value of $h'_{rms}$. A higher $h'_{rms}$ results in a smaller slope $\beta$ (values of $\beta$ are shown in the caption of Fig. 3), whereas other roughness parameters have a limited effect on aging. Since we have previously shown [Fig. 2 (d)] that there is a linear dependence of aging on the real contact area $A$, we hypothesize that the reason why $h'_{rms}$ has a much larger effect on chemical aging than other roughness parameters is also related to the real contact area $A$. According to Eq. (2), among the roughness parameters investigated here only $h'_{rms}$ affects $A$ explicitly. To test our hypothesis, in Figs. 3 (b) and (c), we plot the same data for $\Delta\mu$ as shown in Fig. 3 (a), but this time as a function of the fractional contact area $A/A_0$. Figs. 3 (b) and (c) correspond to a long ($t = 10^3$ s) and a short ($t = 1$ s) hold time, respectively. It is clear that a larger $h'_{rms}$ corresponds to a smaller real contact area $A$. All other parameters, i.e., H, $\zeta$, and $\eta$ have much smaller effects on $A$, which is consistent with Eq. (2).

The results also show that $\Delta\mu$ increases linearly with $A$ for long hold times [Fig. 3 (b)]. This can be attributed to the same mechanism as discussed in Section III A. Specifically, the number of bonds $N$ scales linearly with the real contact area and $\Delta\mu$ is proportional to $N$ by definition. However, in Fig. 3(b) (showing $\Delta\mu$ vs. $A/A_0$) there is a small positive offset on the $\Delta\mu$ axis, which was not found in Fig. 2 (d) (showing $\Delta F$ vs. $A/A_0$). This offset cannot be explained by the area dependence of chemical aging. If $N$ is strictly proportional to the contact area, then $\Delta\mu$ should vanish as the real contact area $A$ goes to zero. Even more surprisingly, $\Delta\mu$ decreases with $A$ for short hold time as shown in Fig. 3 (c), which is the opposite trend to that observed for long hold times. The transition from a monotonously decreasing to a monotonously increasing function of $\Delta\mu$ vs. $A$ occurs around a hold time of $t = 1$-$10$ s, which is the beginning of the regime of logarithmic time dependence.

The above observations can be explained by the Eyring pressure effect [Eq. (3)] on the initial energy barrier to bond formation $E_{b,form}(t = 0)$ as shown in Fig. 4. Fig. 4 (a) shows a local pressure distribution of rough surfaces with different values of $h'_{rms}$. A rougher surface (larger $h'_{rms}$) leads to appearance of larger local contact pressure $P$. Based on the Eyring theory, a larger local pressure will reduce the local energy barrier to bond formation across the interface as shown in Fig. 4 (b). In Fig. 4 we show results of simulations for specific surfaces marked by the open circle (○) in Fig. 3, but the same qualitative trends are observed for other rough surfaces considered in our study. Following the above arguments, rougher surfaces have more reaction sites with reduced $E_{b,form}$, and those highly reactive sites can form bonds within much shorter time period. This is the reason why a rougher surface (larger $h'_{rms}$) can have a higher value of $\Delta\mu$ than a flatter surface (lower $h'_{rms}$) at short hold times as shown in Fig. 3 (c), even though a rougher surface has a lower real contact area $A$. As chemical aging proceeds, highly reactive reaction sites on the rough surfaces will be gradually consumed by bonding with the opposite surface, and the remaining sites will have generally higher energy barrier to bond formation. At a certain point in time, which is around $t = 1$-$10$ s, the effect of the real contact area takes over, resulting in an increase of $\Delta\mu$ with $A/A_0$. Due to the considerable amount of aging that has occurred on rougher surfaces during short hold times, the fitted line to $\Delta\mu$ vs. $A/A_0$ relationship intersects on $\Delta\mu$ axis at long hold time as shown in Fig. 3 (b).

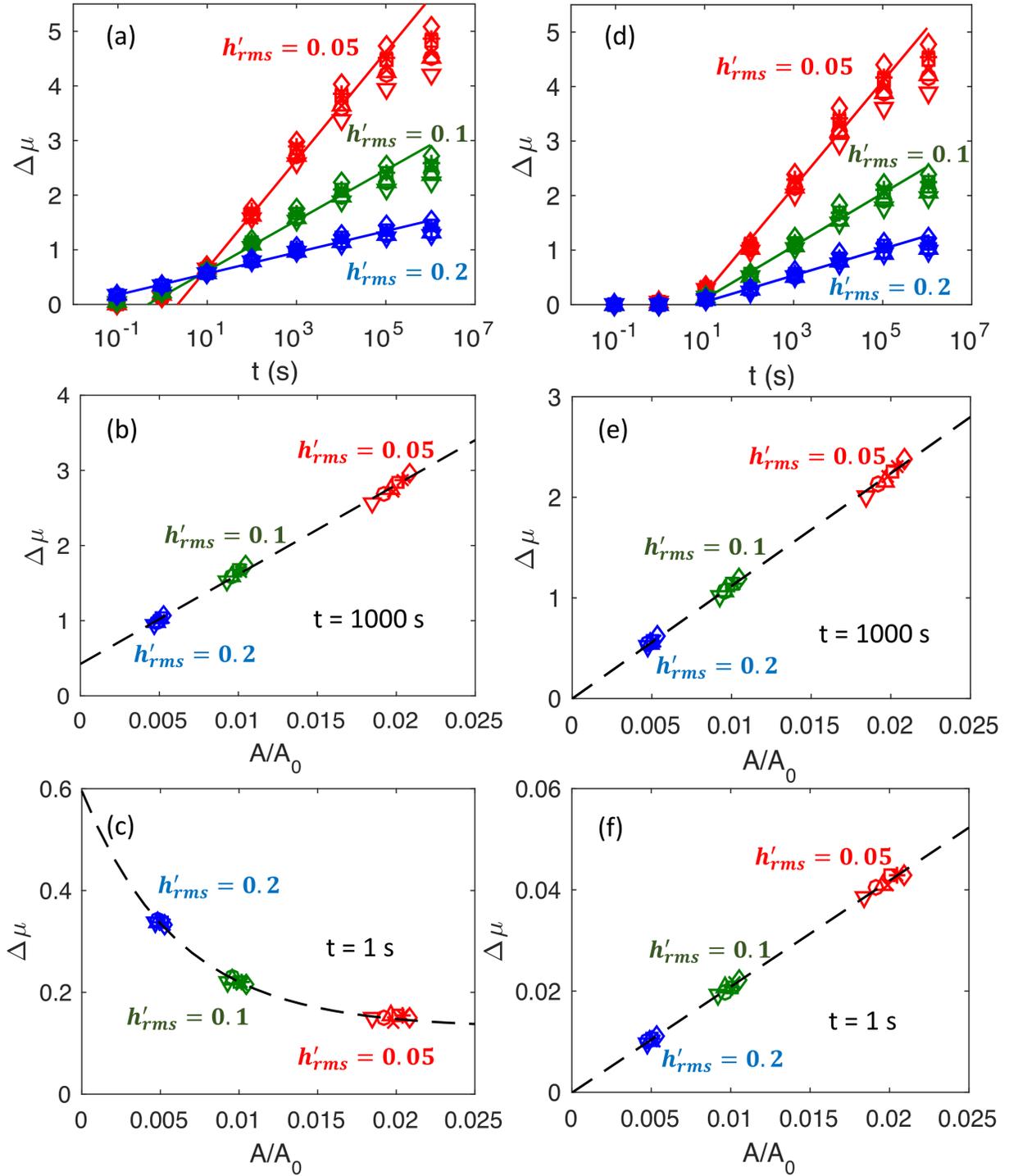

FIG. 3 (color online) Results of kMC simulations for surfaces with different roughness parameters at normal load $F_N = 20$ μN. Red, green, and blue markers represent simulation results for surfaces with $h'_{rms} = 0.05$, 0.1, and 0.2, respectively. The surface roughness parameters corresponding to each marker are shown in Table 1. Red, green, and blue solid lines are linear fits to the simulation results marked by the open circles (○). In (a), the slopes $\beta$ of the fitted red, green, and blue lines are 0.89, 0.45, and 0.2, respectively. In (b), the slopes $\beta$ of the

*fitted red, green, and blue lines are 0.98, 0.49, and 0.24, respectively. For simulations shown in the left column [(a)-(c)], the activation volume $\Delta V$ = 6.4 Å$^3$ (0.04 eV/GPa) [23,52]. For simulations shown in the right column [(d)-(f)], $\Delta V$ = 0.0 Å$^3$. [(a) and (d)] Time dependence of $\Delta\mu$. $\Delta\mu$ as a function of the fraction of the real contact area $A/A_0$ at hold time t = 10$^3$ s [(b) and (e)] and t = 1 s [(c) and (f)]. All results are averaged over 10 different simulations performed for the same conditions (error bars are smaller than the symbol size.)*

*Table 1 Surface roughness parameters corresponding to markers shown in Fig. 3. H is Hurst exponent, $\zeta = q_s/q_L$, $\eta = q_r/q_L$, $N_P$ is the number of contact patches, and $C/A$ is the ratio of circumference (C) to the real contact area (A). Values for $N_P$ and $C/A$ are shown for the case of $h'_{rms} = 0.05$ as a representative example.*

| Marker | H | $\zeta$ | $\eta$ | $N_P$ | $C/A$ (1/nm) |
|---|---|---|---|---|---|
| Circle ○ | 0.8 | 150 | 3 | 160 | 0.44 |
| Square □ | 0.5 | 150 | 3 | 499 | 0.73 |
| Diamond ◊ | 0.3 | 150 | 3 | 1013 | 0.96 |
| Upward-pointing triangle △ | 0.8 | 75 | 3 | 75 | 0.27 |
| Downward-pointing triangle ▽ | 0.8 | 15 | 3 | 14 | 0.11 |
| Asterisk ✶ | 0.8 | 150 | 10 | 435 | 0.64 |
| Cross × | 0.8 | 150 | 1 | 113 | 0.42 |

In order to further demonstrate the existence of the above effect of local contact pressure, we also perform simulations without the Eyring effect, which means that the activation volume $\Delta V$ = 0.0 Å$^3$ in Eq. (3). The simulation results are shown in Figs. 3 (d)-(f). Fig. 3 (d) reveals that the value of $\Delta\mu$ still strongly depends on $h'_{rms}$, but this time $\Delta\mu$ increases proportionally with A at both long and short hold times, as shown in Figs. 3 (e) and 3 (f), respectively. The absence of the pressure effect also causes the delay of the onset of the logarithmic time dependent regime to t ≈ 10 s.

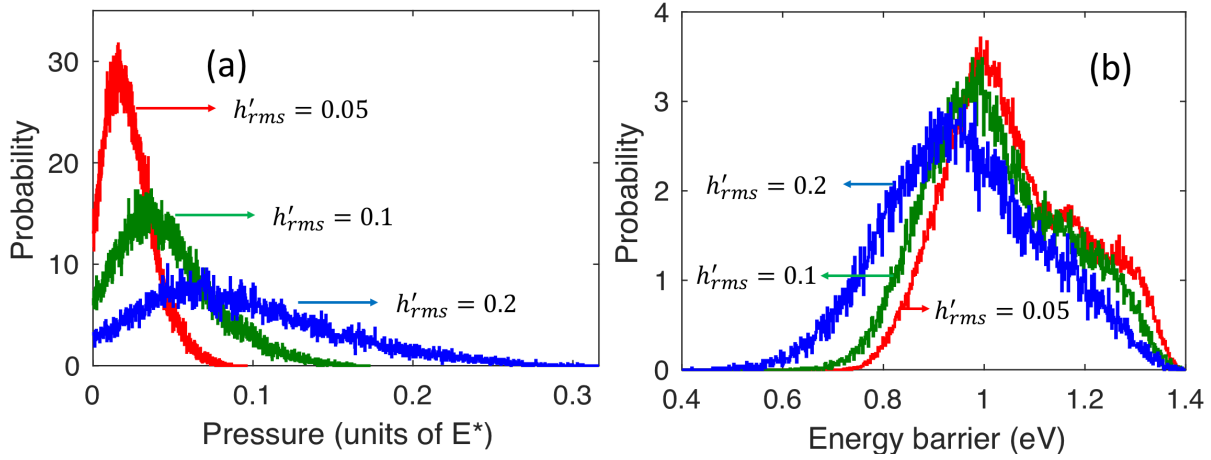

FIG. 4 (color online) Distributions of (a) local contact pressure P and (b) initial local energy barrier to bond formation $E_{b,form}(t=0)$ for rough contacts with different values of $h'_{rms}$ and for $H = 0.8$, $\zeta = 150$, and $\eta = 3$ (marked by open circle (○) symbol in Fig. 3) at normal load $F_N = 20$ μN. Red, green and blue markers correspond to surfaces with $h'_{rms} = 0.05$, 0.1, and 0.2, respectively. Effective Young's modulus $E^* = 35.3$ GPa. All the distributions are normalized so that the areas below the distributions are equal to 1.

Although other roughness parameters, i.e., H, $\zeta$ and $\eta$, have much smaller effect on the contact area and the contact pressure as compared to that of $h'_{rms}$, these parameters can greatly change the surface topography and contact morphology (see Supplemental Material for details [47]). Specifically, higher $\zeta$, higher $\eta$, and lower H promote formation of more uniformly distributed smaller contact patches, whereas lower $\zeta$, lower $\eta$, and higher H lead to contact patches that are larger and clustered. This trend in the number of disconnected contact patches $N_p$ for constant $h'_{rms}$ and $F_N$ is shown in Table 1. The value of $N_p$ further affects the ratio of circumference (C) to the real contact area (A). Here C is defined as the total length of the edges of the boundary elements that grid the surface and that separate contact and non-contact regions. Rough contacts with uniformly distributed contact patches have higher values of $N_p$ and $C/A$. In contrast, those contacts with clustered contact patches have lower $N_p$ and lower $C/A$.

We found that the $C/A$ ratio indeed has a small but clear effect on the reaction kinetics (see Fig. 5). The role of the $C/A$ effect can be explained by the fact that the energy barriers to bond formation $E_{b,form}$ can be affected by previously bonded sites more significantly in the case of fewer and more clustered contacts (a lower $C/A$ ratio). This is due to the elastic interaction between neighboring bonds introduced in our model. In such case, there will be more reaction sites with high value of $E_{b,form}$ and also more reaction sites for which the kinetics of bond formation will be slower than that of bond breaking. It follows that the surface with the same total real contact area but with fewer contact patches should have a lower value of $\Delta\mu$. In Fig. 5, we plot the same data for $\Delta\mu$ as shown in Fig. 3 (e) (i.e., at hold time $t = 10^3$ s and without the Eyring effect), but this time the data is plotted as a function of $C/A$. In addition, $\Delta\mu$ is scaled by the fraction of contact area $A/A_0$ in order to eliminate the effect of the contact area. $\Delta\mu A_0/A$ can be seen as a dimensionless drop in the coefficient of friction, and it also corresponds to the slope of the line shown in Fig. 3 (e). We find that $\Delta\mu$ increases linearly with $C/A$, indicating that the morphology of contact patches does affect chemical aging. However, this contact morphology effect is barely visible in Fig. 3(e), where all the markers fall along a single line. This is because the increase in $\Delta\mu$ with $C/A$ is rather limited, resulting in the slope $\Delta\mu A_0/A$ that is almost constant for surfaces with different roughness. Our analysis shows that although contact morphology does affect chemical aging, this effect is significantly smaller than the influence of the real contact area and the local contact pressure.

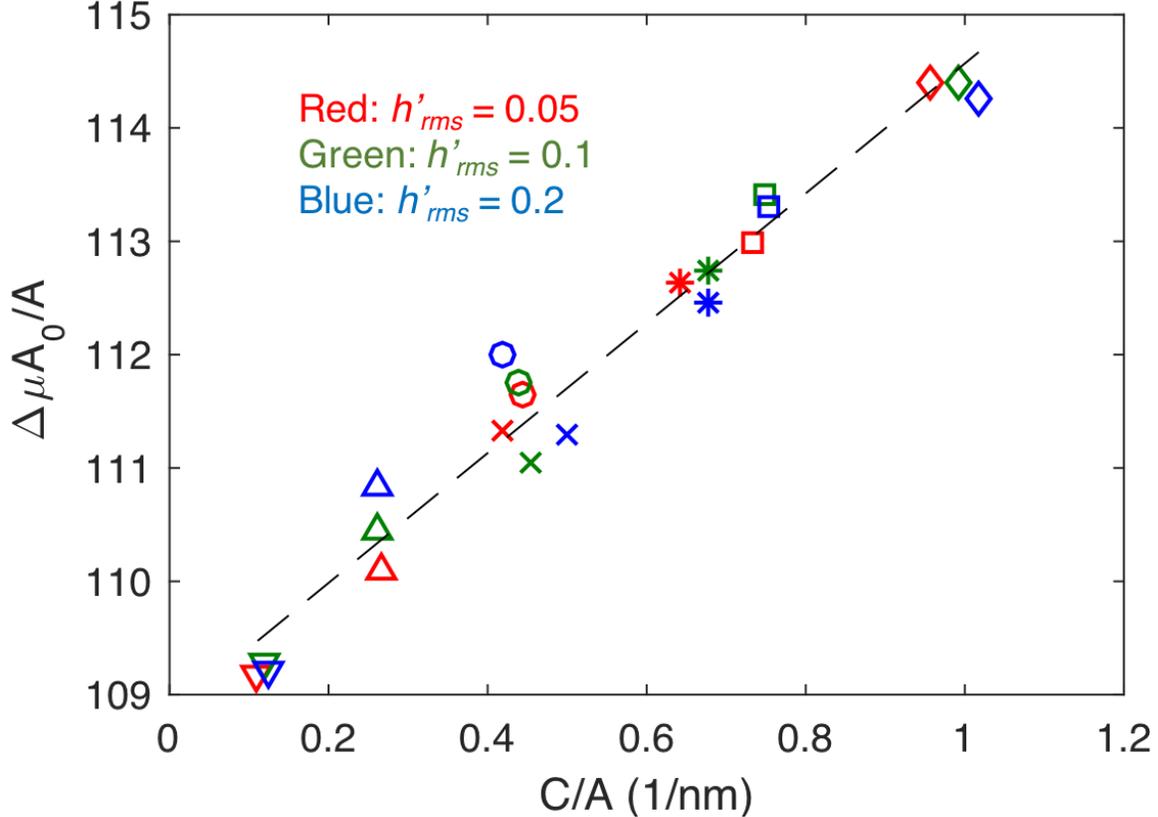

FIG. 5 Drop in the coefficient of friction $\Delta\mu$ vs. the ratio of circumference to real contact area $C/A$ for the surfaces at hold time $t = 10^3$ s. $\Delta\mu$ is scaled by the fraction of the real contact area $A/A_0$. The same data as shown in Fig. 3(e) as a function of $A/A_0$. Marker type represents roughness parameters specified in Table 1.

## IV. DISCUSSION

In our earlier study of chemical aging in single asperity contacts [25] we have found a quantitative agreement between our kMC model and reported AFM measurements of single asperity aging [22]. However, the results obtained here from the same kMC model combined with BEM simulations of randomly rough contacts show a quantitative deviation in the value of $\beta$ (i.e., the rate of increase in $\Delta\mu$) from those reported in macroscopic experiments on silica/quartz interfaces. Specifically, the value of $\beta$ obtained in our rough contact simulation, 0.2-0.9 (see Fig. 3 (a)), is 1-2 orders of magnitude higher than typical experimental values ($\beta \approx 0.01$) obtained for both bare surface contacts [1] and for shearing within a gouge layer [2,20,64–66].

We note that the values of $\beta$ obtained by our model for rough contacts are already much smaller than those measured experimentally in single asperity contacts. The values of $\beta$ for single asperity contacts have not been directly reported, but they can be inferred from AFM experiments, [2] i.e., $\beta \approx 15$ for $F_N$ = 23 nN and, $\beta \approx 1.5$ for $F_N$ = 393 nN. Our multiscale model predicts that there is 1-2 orders of magnitude reduction in $\beta$ between single-asperity and rough contacts, mainly due to the large reduction in the fraction of the real contact area. However, this reduction is not yet sufficient to reach a quantitative agreement with macroscopic experiments.

It is interesting to ask what might be reasons for the deviations in $\beta$ between simulations and experiments on rough contacts. Possible phenomena that could be responsible for this deviation include: 1) partial hydroxylation of surfaces (instead of a complete hydroxylation assumed in our model), 2) higher values of $h'_{rms}$ than those assumed in our model, 3) a so-called detachment front propagation (explained below) for bare surface contacts, and 4) relative contributions from sliding and rotation of granular particles for contacts within a gouge layer.

Regarding hydroxylation, $SiO_2$ surfaces used in macroscopic experiments of aging may have a lower density of OH group on the surface than what is assumed in our study. We assumed a value corresponding to a fully hydroxylated surface of $\rho_{OH}$ = 4.9 OH/nm$^2$ based on Ref. [42]. However, the actual surface density of hydroxyl groups may depend on how the samples are prepared [67,68]. Partial hydroxylation would result in the presence of non-reactive surface Si-O-Si groups that cannot easily form interfacial siloxane bonds. A few experiments suggested [69,70] that, unless the surfaces of the samples are treated in a special way, the density of OH groups on amorphous $SiO_2$ fracture surfaces saturates at $\rho_{OH}$ = 2.6 OH/nm$^2$, at partial pressures of $H_2O$ up to 1333.2 Pa. The silica surfaces used in single asperity AFM experiments of aging (where the rate of aging is high and comparable to that predicted by our simulations) have been treated explicitly with piranha solution, which makes the surfaces strongly hydrophilic [22]. This is in contrast to surfaces used in many macroscopic aging experiments where prior to friction measurements the sample were only oven dried [1] or held under humid environment for certain time [20] without further cleaning treatment. Recent theoretical studies [71–73] also indicate that the density of OH group on silica/quartz surface is sensitive to the temperature and partial pressure of $H_2O$, and thus the surface chemistry of the experimental sample should be carefully investigated before the models can be quantitatively predictive.

The second possible reason for the quantitative discrepancy in $\beta$ is the difference in the value of $h'_{rms}$ encountered on actual surfaces and the surfaces considered in our model. If the actual surface has higher value of $h'_{rms}$, $\beta$ would decrease, consistently with the trend shown in Fig. 3. Unfortunately, measurements of $h'_{rms}$ are not straightforward in practice. The value of $h'_{rms}$ greatly depends on the shortest wavelength roughness components, or the large wavevector cut-off $q_s$, and measurement of this value requires high-resolution scanning of surfaces. The experimental values of $h'_{rms}$ are not widely reported, but a natural fault surface measured in recently reported experiments [74] are found to have the roughness of $h'_{rms}$ = 0.26 for wavelengths scanned from 10 μm to 60 nm. This value is already higher than the highest rms slope of $h'_{rms}$ = 0.2 we used in Fig. 3. Furthermore, the extrapolated value of $h'_{rms}$ of this fault surface down to the atomic length scale leads to $h'_{rms} \approx 1.0$ [75]. For such a high rms slope, surfaces in contact are subject to some amount of plastic deformation. It was previously shown that elasto-plastic deformation in contacts with yield strength $\sigma_y$ = 0.01$E^*$ can lead to an increase in the contact area by approximately a factor of two as compared to purely elastic contacts [76]. Since silica has a higher yield strength of $\sigma_y \sim 0.1E^*$ [77], therefore the increase in the contact area due to the plastic deformation for silica could be significantly lower than a factor of 2. If the surface has $h'_{rms} = 1.0$ and we take the upper limit on the amount of plastic deformation, the contact area could be roughly estimated as $2 \times 0.2 / 1.0 = 0.4$ times of the contact area of the surfaces with $h'_{rms} = 0.2$ [contact area is inversely linear with $h'_{rms}$ according to Eq. (2)]. It means, the value of $\beta$ would decrease from what is predicted by our simulations, even with plastic deformation as long as $h'_{rms}$ is high enough.

Another contribution to the discrepancies in $\beta$ measured experimentally and predicted by our simulations could be a phenomenon referred to as the detachment front propagation across the interface. It has been observed experimentally that for macroscopic contacts, the detachment of the contact does not always take place simultaneously across the entire contact interface. Instead the contact area is first reduced gradually by propagation of the detachment fronts [78–80] and only after a certain time sliding of the entire contact takes place. Experiments [81] and simulations [82] showed that the detachment front phenomenon in macroscopic contacts could lead to large variation in the static friction depending on the loading configuration. For macroscopic experiments on bare surface contacts, two blocks of the material are slid against each other by applying a force at the edge of the block [1,83], rather than by applying a uniform shear force across the interface. In this case, the maximum reduction in the static friction force due to the detachment front effect is ~85 % [82]. This analysis suggests that our simulation results of $\beta$ shown in Fig. 3 might be overestimated. In experiments on gouge layer contacts, the samples are sheared in a biaxial deformation apparatus [13,65], and the external shear stress is applied uniformly over the area of the interface. Thus, the effect of the detachment front discussed above may not be applicable to gouge layer experiments. However, in these experiments the friction coefficient could be lower than predicted by our model because of the particle rotation within the gouge layer. Earlier simulation using the discrete element method (DEM) [84] have shown that friction within the gouge layer results from both sliding and rotation of granular particles. Additionally, it was shown that the rotational component may lead to limited increase in overall friction coefficient $\mu_f$, even if the interparticle friction coefficient $\mu_p$ increases (e.g., $\mu_f$ increases only from 0.2 to ~0.3 when $\mu_p$ is changed from 0.1 to 0.5). The reason is that when the local shear resistance $\mu_p$ increases, it becomes more difficult for the particles to slide past each other and they begin to rotate instead. Since this effect is not included in our simulations, it may explain the lower values of $\beta$ measured in typical gouge layer aging experiments.

In order to estimate the effect of the aforementioned phenomena on $\Delta\mu$, we performed additional kMC simulations where we assumed a bare surface contact, which is only partially hydroxylated ($\rho_{OH} = 0.5 \times 4.9$ OH/nm$^2$). The results for roughness parameters $h'_{rms} = 0.4$, $H = 0.8$, $\zeta = 150$, and $\eta = 3$ are shown in Fig. 6. The value of $\Delta\mu$ obtained from kMC simulation is multiplied by 0.15 to account for the 85% decrease in friction force due propagation of detachment fronts [82]. The activation volume $\Delta V$ is set to 0.04 Å$^3$. When all these factors are considered, the value of $\beta = 0.0087$ is obtained, which is on the same order of magnitude as the results of macroscopic experiments. Gouge layer contacts are not examined here, because the value of $\mu_f$ (0.2-0.3) obtained from the DEM simulations is somewhat smaller than the typical experimental value (~0.65 [20,65]), and we do not know the exact scaling factor between $\mu_f$ and $\mu_p$. However, it can be expected that the value of $\beta$ also decreases, which is the desired trend, at least qualitatively.

A more quantitative test of the model would require a well characterized macroscopic experiments, where both the surface chemistry and surface roughness are analyzed in detail.

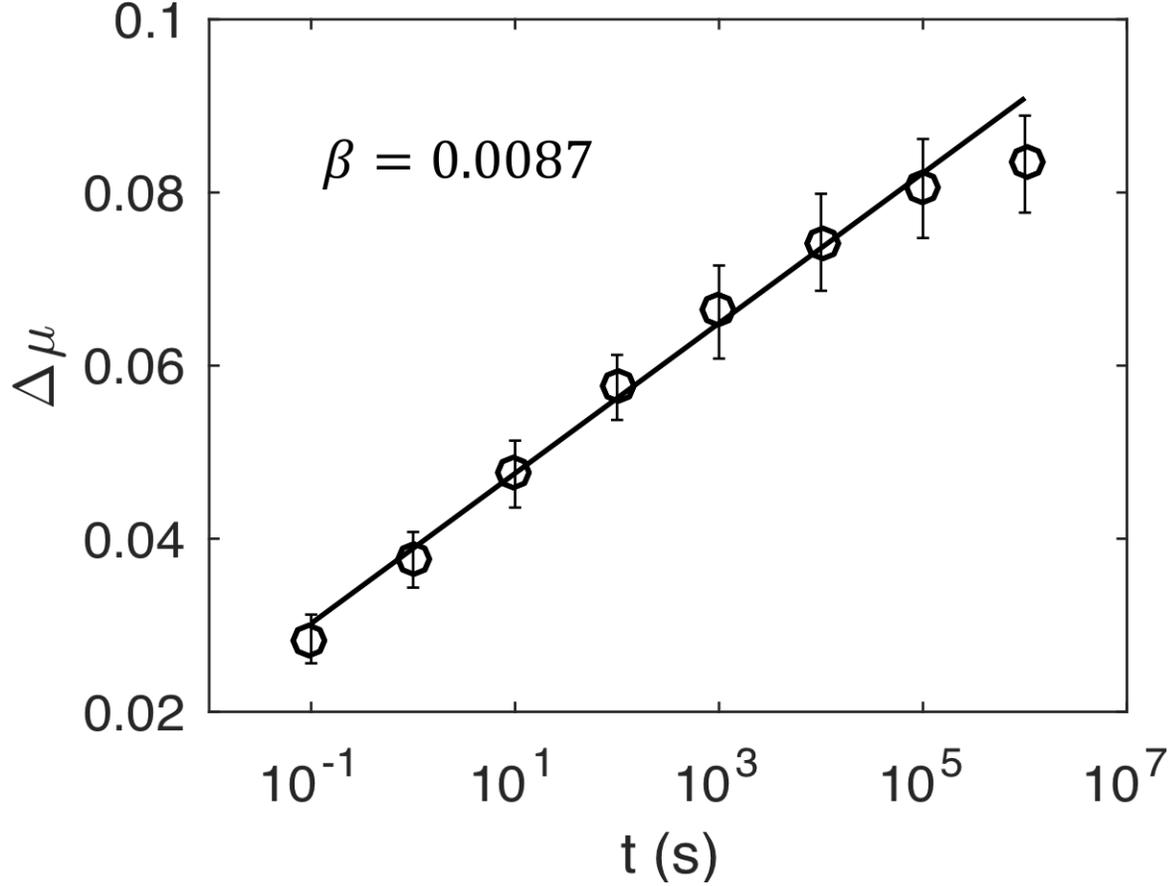

FIG. 6 Time dependence of the drop in the coefficient of friction $\Delta\mu$ on partially hydroxylated surface ($\rho_{OH} = 0.5 \times 4.9$ OH/nm$^2$) with roughness parameters $h'_{rms} = 0.4$, $H = 0.8$, $\zeta = 150$, and $\eta = 3$. The values of $\Delta\mu$ obtained from kMC simulations are multiplied by 0.15 to include the detachment front effect. The results are averaged over 10 different simulations performed for the same conditions.

## V. CONCLUSION

We combined kMC model of chemical aging with BEM simulations for contact mechanics to study chemical aging of large-scale $SiO_2$-$SiO_2$ non-adhesive randomly rough contacts. We found that the regime of logarithmic time-dependence and linear load-dependence of chemical aging observed for single asperity contacts also exist in randomly rough contacts. Contact aging (quantified by a drop either in the friction force or in the friction coefficient) depends linearly on the applied load. This functional relationship holds as long as the real contact area increases linearly with the applied load.

Our multiscale kMC/BEM model with different surface roughness parameters showed that the chemical aging strongly depends on the rms slope $h'_{rms}$ of the surface. This is because $h'_{rms}$ not only affects the real contact area $A$, but also alters the local contact pressure distribution which further changes the initial distribution of energy barriers to bond formation $E_{b,form}(t = 0)$. In addition, the contact morphology, which is controlled by other roughness parameters, i.e., H, $\zeta$ and $\eta$, does affect the chemical aging behavior through the elastic interaction between neighboring reaction sites for siloxane bond formation. However, this effect is much smaller than the effects of the real contact area and of the local pressure. We propose several additional phenomena that may need to be included in modeling of aging on rough macroscopic surfaces in order to explain the remaining quantitative discrepancy in the estimate of aging rate (i.e., $\beta$) obtained in our simulations and in experiments on rough contacts.

**ACKNOWLEDGMENT**

The authors gratefully acknowledge financial support from the National Science Foundation (grant #1549153) and the Deutsche Forschungsgemeinschaft (grant PA 2023/2). We thank R. Carpick, K. Tian, D. Goldsby and C. Thom from University of Pennsylvania, and Yun Liu from Apple Inc. greatly for helpful discussions and suggestions.